\documentclass{agujournal2019}
\usepackage{url} 
\usepackage{lineno}
\usepackage[inline]{trackchanges} 
\usepackage{soul}

\usepackage{booktabs}
\usepackage{mhchem}
\usepackage{ragged2e}             

\usepackage{pdfpages}

\draftfalse

\journalname{Geophysical Research Letters}

\begin{document}

\title{Resilience of Snowball Earth to Stochastic Events}

%
%




\authors{Guillaume Chaverot\affil{1,2,3}, Andrea Zorzi\affil{4}, Xuesong Ding\affil{5}\thanks{Now at Bureau of Economic Geology, The University of Texas at Austin, Austin, TX, US}, Jonathan Itcovitz\affil{6,7}, Bowen Fan\affil{8}, Siddharth Bhatnagar\affil{1,3,9}, Aoshuang Ji\affil{10}, Robert J. Graham\affil{8}, Tushar Mittal\affil{10}}

\affiliation{1}{Observatoire Astronomique de l'Universit\'e de Gen\`eve, Chemin Pegasi 51, CH-1290 Versoix, Switzerland}
\affiliation{2}{Univ. Grenoble Alpes, CNRS, IPAG, 38000 Grenoble, France}
\affiliation{3}{Life in the Universe Center, Geneva, Switzerland}
\affiliation{4}{Department of Earth and Planetary Sciences, Stanford University, Stanford CA, US}
\affiliation{5}{Department of Earth, Planetary, and Space Sciences, University of California, Los Angeles, CA, US}
\affiliation{6}{Institute of Astronomy, University of Cambridge, Madingley Road, Cambridge, UK}
\affiliation{7}{Department of Civil and Environmental Engineering, Imperial College London, London, UK}
\affiliation{8}{Department of the Geophysical Sciences, University of Chicago, Chicago, IL, USA}
\affiliation{9}{Group of Applied Physics and Institute for Environmental Sciences, University of Geneva, Bd. Carl-Vogt 66, CH-1205 Geneva, Switzerland}
\affiliation{10}{Department of Geosciences, Pennsylvania State University, University Park, PA, US}




\correspondingauthor{Guillaume Chaverot}{guillaume.chaverot@univ-grenoble-alpes.fr}




\begin{keypoints} 
\item We use an Energy Balance Model (EBM) to simulate the response of Snowball climate after stochastic events.
\item Impact simulations and estimates of ash dispersion are used to inform initial conditions.
\item Earth's Snowball states seem resilient to termination by stochastic events within our modeling framework.
\end{keypoints}

\begin{abstract}
\justifying\noindent
Earth went through at least two periods of global glaciation (i.e., ``Snowball Earth'' states) during the Neoproterozoic, the shortest of which (the Marinoan) may not have lasted sufficiently long for its termination to be explained by the gradual volcanic build-up of greenhouse gases in the atmosphere. Large asteroid impacts and supervolcanic eruptions have been suggested as stochastic geological events that could cause a sudden end to global glaciation via a runaway melting process. Here, we employ an energy balance climate model to simulate the evolution of Snowball Earth's surface temperature after such events.
We find that even a large impactor (diameters of $d \sim 100\,\mathrm{km}$) and the supervolcanic Toba eruption ($74\,\mathrm{kyr}$ ago), are insufficient to terminate a Snowball state unless background CO$_2$ has already been driven to high levels by long-term outgassing. We suggest, according to our modeling framework, that Earth's Snowball states would have been resilient to termination by stochastic events.
\end{abstract}

\section*{Plain Language Summary} 
\justifying\noindent
The terminations of Earth's longest periods of global glaciation are commonly understood to have occurred due to the gradual build-up of greenhouse gases in the atmosphere from volcanism. However, the sudden ends of Earth's shorter global glaciation periods likely cannot be explained by the same mechanisms. Large asteroid impacts and supervolcanic eruptions have been suggested as geophysical phenomena that could cause abrupt ends to global glaciation periods. Here, we model the evolution of the planet's surface temperature in the aftermath of such events. Impacts and eruptions open up gaps in the global ice sheet, and also partially cover the ice in far-spreading dust and ash, both of which increase the amount of solar radiation that is absorbed by the planet comparing to the highly reflective surface of ice and snow. Greater absorption of radiation leads to higher surface temperatures, which increases ice melting, and generates a feedback loop that can melt the entire planet surface. 
However, we find that the scales of impact or eruption required to produce global melting are too great to have likely occurred at the times of Earth's global glaciations. Other mechanisms must, therefore, be explored to explain Earth's short glaciation periods.

\section{Introduction}\label{sec:Introduction}
\noindent
The Snowball Earth hypothesis \cite{Kirschvink1992, Hoffman1998,hoffman_snowball_2017} proposes global surface coverage of Earth in a thick ice layer. In contrast, the most recent glacial episode, the Last Glacial Period, witnessed only partial ice coverage extending down from the poles but not reaching equatorial regions \cite<e.g.,>{batchelor2019}. Snowball events have occurred on at least two occasions during the Neoproterozoic era ($1000-538.8 \,\mathrm{Ma}$). In particular, there is good evidence for a snowball event during the Cryogenian period ($720-635 \,\mathrm{Ma}$), with other potential Snowball states also being suggested \cite<e.g.,>{chumakov2009baykonurian}. Evidence for Snowball Earth's existence comes in many forms, including paleomagnetic data in Australia \cite{Harland1964,hambrey1981earth}, Australian cap carbonates \cite{kennedy1996stratigraphy}, and extreme carbon isotope excursions in Norway and Greenland \cite{knoll1986secular}. Geologic measurements have been complemented by modeling efforts \cite<e.g.,>{Caldeira1992} to understand both the origins and demise of the Snowball state.

Under Earth's current insolation, a Snowball state is one of three possible stable climate regimes for Earth, alongside a temperate state and a hot steam atmosphere state \cite{shields_spectrum-driven_2014,Turbet2021}.
However, as shown in \citeA{brunetti_co-existing_2019}, the number and nature of these stable climate regimes can vary depending on feedbacks considered within the model.
The Snowball state is stable against moderate variations in insolation through the high albedo of ice, which acts to prevent increases in global temperature \cite{shields_spectrum-driven_2014, hoffman_snowball_2017}. This creates a bi-stability between the temperate Earth and Snowball Earth through the difference of albedo between ice and liquid water \cite{ghil1994cryothermodynamics}. It is, therefore, impossible to deglaciate Snowball Earth under such conditions without other climate forcings. 
One such forcing that is commonly invoked is the build-up of atmospheric \ce{CO2} \cite{walker1981negative,menou2015climate}. Magmatic volatile emissions from volcanism, coupled with the absence of \ce{CO2} sinks of Earth's temperate-state carbon cycle (e.g., silicate weathering being prevented by the covering of ice) leads to a strong greenhouse effect that warms the planet, triggering the Snowball termination \cite{Kirschvink1992,Caldeira1992}. However, the time required to accumulate sufficient \ce{CO2} to trigger melting \cite<$4-30 \,\mathrm{Myr}$, >{Hoffman1998} is too long to account for the short Marinoan glaciation \cite{Rooney2015}. Suggested alternatives include impact events \cite{Kring2003} and massive volcanic eruptions \cite{Zhongwu2021}, which we hereafter refer to as stochastic events due to the nature of their occurrences. However, quantitative estimates of climate after stochastic events remain relatively unexplored.

Impacts of large asteroids can generate global climate effects \cite<e.g.,>{Brugger2017,Pierazzo1998,Artemieva2017,turbet_environmental_2020}. Simulations of such impacts have estimated the masses of water vapor injected into the atmosphere by these events, but the climatic implications have not been investigated  \cite{koeberl2019asteroid} or have been inconclusive due to uncertainties on the radiative effect of water vapor \cite{Erickson2020}. Nonetheless, the temporal coincidence between the youngest Paleoproterozoic glacial deposits and the Yarrabubba event has sparked interest in the role of impacts in climate evolution \cite{Erickson2020}. Further, given their large kinetic energy, impacts can melt the ice table covering the planet surface and vaporize rocks from the underlying crust. Therefore, among other effects, impacts can (1) increase the planet's surface temperature by hundreds of degrees for large distances from the impact site, and (2) deposit ejecta material onto the icy surface far from the impact site, affecting the planet's surface albedo.

Supervolcanic eruptions are also suggested candidates to escape the Snowball Earth state. Such events are defined by either their magnitude ($M\geq8$, proportional to the erupted mass) or their ejecta volume \cite<$1,000 \,\ \mathrm{km^3}$,>{desilva2022}, and are thousands of times larger than typical volcanic eruptions \cite<e.g.,>{Hansen1978}. The largest known super-eruption is Toba ($M=9.1$), which is recorded in the 74,000-year-old Youngest Toba Tuff \cite<YTT,>{Rose1990}. The ash ejected by the event covered $\sim$ 40 million $\mathrm{km^2}$ of Earth's surface at a depth greater than 5 mm \cite{Costa2014}. Toba-like super-eruptions are expected to happen $1-2$ times per million years \cite{Mason2004, cisneros_de_leon_synchronous_2022}. Supervolcanic eruptions have significant global climate effects \cite<e.g.,>{Hansen1978}. Large volumes of volcanic gases are injected into the atmosphere, such as (1) \ce{CO2}, which acts as a classical greenhouse gas, (2) \ce{SO2}, which acts as a solar-reflecting coolant, and (3) \ce{H2O}, which can act as a greenhouse gas, but can also form radiative-active clouds. Further, eruptions deposit ash and dust on the icy surface, reducing the albedo in comparison to clean ice and hence increasing the amount of sunlight absorbed by the planet's surface. Although often cited as a possible cause of the onset of global glaciation \cite{macdonald_initiation_2017}, there is no detailed exploration of the viability of this scenario.

In this study, we quantify the heating effects of stochastic events on a globally glaciated Earth, testing the potential of such events to generate a runaway melting process that would lead to the termination of the Snowball state. Through impact simulations and ash-/ejecta-dispersal modeling (Sections \ref{sect:impact} \& \ref{sec:Supervolanoes}), we quantify the changes in albedo and surface temperature produced by such events. Using a 1-D Energy Balance Model (EBM; Section \ref{sec:EBM}), we then analyze the radiative perturbation caused by such temperature and albedo changes, following the evolution of the post-event surface temperature profile and ice coverage. By varying the scale of our stochastic events (e.g., impactor size, magnitude of supervolcanic eruption), we assess the capabilities of such events to trigger global deglaciation (Section~\ref{sec:results}). We discuss the likelihood of occurrence of our event scales during the Neoproterozoic, and thus the likelihood of deglaciation by stochastic events, in Section~\ref{sec:discussion}.

\section{Method} \label{sec:methods}
\subsection{Climate modeling}\label{sec:EBM}
\citeA{budyko1969} and \citeA{sellers1969}'s independent seminal works on the Earth's climate established the usefulness and robustness of 1-D EBMs. In a latitudinal EBM, the prognostic surface temperature is zonally averaged, thus leaving only the meridional dimension. 
This gives a remarkably good first-order estimate to solving the climate of a rocky planet \cite{north1979,spiegel_habitable_2008,dressing_habitable_2010}. The 1D-nature of the planet is further exemplified by parameterizing the meridional atmospheric heat transport as temperature-driven diffusion.
Because bi-stability between the temperate Earth and Snowball Earth is a radiative balance process that exists through the difference of albedo between ice and liquid water \cite{ghil1994cryothermodynamics}, we use an EBM focused on radiative balance \cite{williams1998,spiegel_habitable_2008,dressing_habitable_2010} to model the evolution of snowball deglaciation. The corresponding temperate and snowball temperature profiles are shown in Supplementary Figure S2. As 1-D EBMs are not longitudinally defined, we developed a method to account for spatially-located events, correcting input calculations from impact simulations and estimations of ash-dispersion from eruptions (see Supplementary Text S4). 

For our simulations, we assume a planet following a circular orbit at 1 AU around a Sun-like star, with a modern Earth solar constant. We fix the planetary obliquity to 0° and also assume zero eccentricity. Land and oceans are uniformly distributed across the planet such that each latitude cell consists of a 70:30 modern Earth ocean-to-land ratio. Heat capacity and albedo parameterizations between ice and liquid water are smooth transition functions taken from \citeA{williams_habitable_1997} and \citeA{spiegel_habitable_2008} (see Supplementary Text S4). To constrain our study within a simple framework, we neglect the effects of aerosols and ice-melting latent heat: potential caveats from our assumptions are discussed in Section~\ref{sec:discussion}. However, the neglected processes (e.g., dust enrichment of the stratosphere due to impacts, \ce{SO2} production of volcanism) tend to cool the atmosphere, making deglaciation more difficult to achieve. Even in these conditions, a significant warming of the planet is hardly reachable as shown in Section~\ref{sec:results}. Nonetheless, we account for variable relative humidity (RH), cloud forcing \cite{abbot_resolved_2014} and the potential addition of \ce{CO2} released by volcanic eruptions. We also evaluate the influence of the location of stochastic events by modeling eruptions and impacts at the equator and at 40°~N. The effect of impacts is modeled by modifying the initial temperature distribution as per hydrocode simulations (see Section~\ref{sect:impact}). Eruptions do not modify the temperature distribution, but affect the surface albedo due to ash-dispersion.  

\subsection{Impact simulations}\label{sect:impact}
We use the shock physics code iSALE \cite{Collins2016} to model asteroidal impacts on a Snowball Earth state and produce 1-D thermal profiles for the surface of the planet in the aftermath of the impact. We additionally analyze the spread of the impact ejecta curtain in order to estimate the reduction in surface albedo associated with covering the ice with ejecta material.

We initialize simulations based on estimates of Snowball Earth's surface state \cite{Hoffman1998}. A $5\;\mathrm{km}$ thick global ice sheet thus sits at the surface of the target planet, with a rocky crust, mantle, and iron core sitting below. An average surface temperature of $227\;\mathrm{K}$ is prescribed for the pre-impact temperature profile, matching the steady-state temperature found by the EBM in the absence of a stochastic climate forcing event. Impactors consist entirely of dunite, matching closely the thermodynamic behaviors of the chondritic material that are expected from such objects \cite{Benz1989}. We run simulations for impactors with diameters between $40-100 \;\mathrm{km}$ to assess the effects of impact scale. The impact velocity is constant for all impacts at $1.5 \times$ mutual escape velocity ($v_\text{esc}$), representing the most likely impact velocity \cite{CHYBA1991, le2011nonuniform}, and we consider only head-on collisions, which is an approximation accounting for the small mass of the impactors relative to the Earth (more informations about the impact setup are provided in Supplementary Text S2). 

Simulations are run until the motion in the mantle region excavated by the impact settles out (i.e., the mantle has rebounded and subsequent oscillatory motions have terminated). A 1-D surface temperature profile is then extracted, and used as the initial perturbation of the EBM. Because we use a 1-D latitudinal EBM, the initial temperature at the impact site results from the average of the high-temperature impact site longitudes and the low-temperature unperturbed longitudes (see Supplementary Text S4 for more details). As a result, the initial temperature in the EBM (zonal mean temperature, $200-350 \,\mathrm{K}$) is lower than the temperature at the impact site (local temperature, $>1000 \,\mathrm{K}$). 
We find that this approach slightly facilitates deglaciation by marginally overestimating the surface albedo ($<1 \,\%$).

An albedo profile is similarly determined, accounting for the exposure of crustal material after melting of the ice layer and the deposition of material from the impact ejecta curtain on top the ice table outside of the impact crater. 
We assume that ejecta reaches distances up to $5 \times$ the crater radius \cite<e.g.,>{RICHARDSON2005}, corresponding to distances of $2250 \;\mathrm{km}$, $3850 \;\mathrm{km}$, $3900 \;\mathrm{km}$ and $5000 \;\mathrm{km}$ from the impact site for the $40 \;\mathrm{km}$, $60 \;\mathrm{km}$, $80 \;\mathrm{km}$ and $100 \;\mathrm{km}$ impactors, respectively. The albedo for ejecta-covered areas is set to $0.2$, similar to that of cryoconite \cite{HOTALING2021}. A step-like transition is then assumed between dark ejecta and ice albedo ($0.7$). Finally, we re-scale the temperature increase with the surface area that is affected by the impact and the size of latitudinal band, assuming the impact energy is conserved during the zonal transport. The temperature rise is submitted as the input to the EBM.

\subsection{Supervolcanic eruptions}\label{sec:Supervolanoes}

We implement the formalism of \citeA{pyle1989thickness} to model the variation of volcanic ash thickness ($d$) with distance from the eruption center. The exponential relationship between $d$ and the isopach area $A_\mathrm{iso}$ is given by:

\begin{equation}
    d = d_0 e^{-k A_\mathrm{iso}^{1/2}}
    \label{eq:Ash Spread}
\end{equation}
where $d_0$ is the ash maximum thickness (units of $\mathrm{m}$) and $k$ is the rate of ash thinning with isopach area, also in $\mathrm{m}$ \cite{pyle1989thickness}. We find $k$ using ($d$, $A_\mathrm{iso}$) values of volcanic ash fall from Toba, reconstructed through several tens of thickness measurements of the YTT tephra deposit \cite{Costa2014} and digitized with QGIS \cite{QGIS_software}.

The ash deposition region is divided into fully- ($d >0.014 \,\mathrm{mm}$) and partially-covered ($0.002<d<0.014 \,\mathrm{mm}$), the latter known as dusty ice \cite{le2010toward}. Assuming an elliptical geometry for ash-dispersal, the ash deposition area is converted in latitude and longitude. For Toba, the fully-covered area extends for $\pm 43^\circ$ in latitude and $\pm 86^\circ$ in longitude; the dusty ice area spans a wider range between $\pm 50^\circ$ in latitude and $\pm 100^\circ$ in longitude. At visible wavelengths, coarse dust has an albedo of $0.18$ \cite{warren1982optical}. Thus, assuming a clean ice albedo of $0.7$, we assign the dusty ice albedo as changing linearly between clean ice and dust (Supplementary Figure S3). Such values are consistent with measurements of bare ice mean albedo \cite{warren2002snowball}. Fine dust correspond to higher albedo values, thus making deglaciation harder \cite{flanner_snicar-adv3_2021}.

\begin{figure}[t!]
    \centering
    \includegraphics[width=0.92\textwidth]{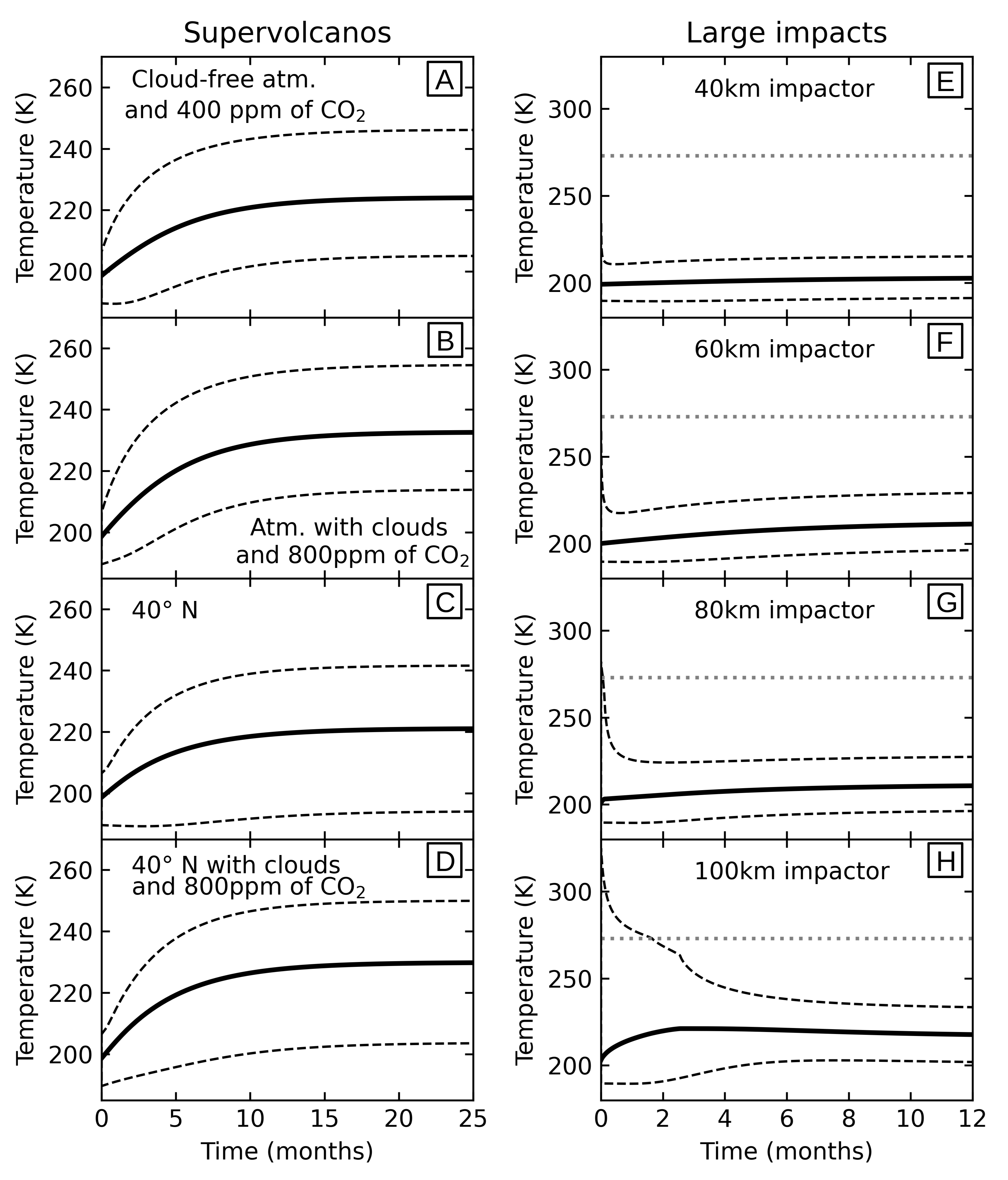}
    \caption{Temporal evolution of surface temperature after a Toba-like eruption (panel A-D) and impacts of varying size (panel E-H). For the eruptions, each panel corresponds to a different set of initial conditions, highlighting the differences that such effects produce under the same climate forcing event. For the impacts, all panels show the same cloud-free atmosphere with $400\;\mathrm{ppm}$ \ce{CO2} of (A). The global-mean temperature (solid lines), minimum and maximum temperatures (dashed lines), and the melting temperature of the surface ice (dotted lines) are all shown.} 
    \label{fig:temp_evol}
    \vspace{-5mm}
\end{figure}

\section{Results} \label{sec:results}
\noindent
The EBM is able to provide the temporal evolution of surface temperature for up to several years after the moment of climate forcing. We find, however, that temperatures tend to converge within around two years (e.g., Figure~\ref{fig:temp_evol}). Geospatial variation in temperature, due to the regional nature of the stochastic events, can be well represented by recording global temperature minima, maxima, and mean averages (Figure~\ref{fig:temp_evol}). 

We first analyzed the effects of a Toba-like eruption (Figure~\ref{fig:temp_evol}A-D) for a cloud-free atmosphere similar to the present-day Earth atmosphere, with variable water vapor, $1 \;\mathrm{bar}$ of nitrogen and $400 \;\mathrm{ppm}$ of \ce{CO2}, and with the eruption happening at the equator (Figure~\ref{fig:temp_evol}A). Additional simulations then included cloud radiative effects \cite<$14 \,\mathrm{W \ m^{-2}}$,>{abbot_resolved_2014} and increased atmospheric \ce{CO2} at $800 \,\mathrm{ppm}$ (Figure~\ref{fig:temp_evol}B) in order to test the robustness of the result to initial atmospheric conditions, and to account for greenhouse gases released by the eruption. Finally, we repeated these set simulations with the climate forcing occurring at $40^\circ \,\mathrm{N}$ (Figure~\ref{fig:temp_evol}C). In none of these simulations does a Toba-like eruption change surface temperatures sufficiently to escape the Snowball state. In the best-case scenario for deglaciation (i.e., accounting for the cloud-warming effect and with $800 \,\mathrm{ppm}$ of \ce{CO2}), the maximum equatorial temperature is only $255 \,\mathrm{K}$ after relaxation from forcing. Other effects were further tested, but were found to contribute minorly to results within realistic bounds, including: the addition of humidity due to volcanic moisture, weaker atmospheric circulation, different ice albedo, and greater effective stellar insolation (see Supplementary Text S5 and Figure S1). 

In the case of our impactors, the largest ($d = 100 \;\mathrm{km}$) generates a transient water belt (note this may not be a true water belt but rather a function of the EBM's spatial structure), with maximum extents of $4.5^\circ \;\mathrm{N}$ and $4.5^\circ \;\mathrm{S}$, and lasting for around two months (Figure~\ref{fig:temp_evol}H). During this time, the surface temperature of the planet decreases due to thermal emission. Once the freezing point is reached, the change of albedo induces a rapid runaway re-glaciation of the transient water belt. We thus find that even the $100 \;\mathrm{km}$ impactor cannot melt a surface large enough to break the ice albedo feedback, which prevents the Snowball Earth deglaciation. The lower resultant temperatures and reduced geospatial domain of our smaller impactors ($d = 40-80 \;\mathrm{km}$) results in no production of a water belt and indeed limited melting of the planet's surface (Figure~\ref{fig:temp_evol}E-G), with the same result of no deglaciation.
Impactors larger than $100 \;\mathrm{km}$ will melt more ice and will generate greater temperature perturbations. However, such impacts would also melt part of the planet crust and/or mantle, for which there is little evidence in the geologic record, and such massive impactors are highly unlikely by the time of the Neoproterozoic (Section~\ref{sec:discussion}). A simple analytical calculation tends to show that the energy required to deglaciate could only be delivered by an impactor that is larger than 100~km (Text~S1).

\begin{figure}[t!]
    \centering
    \includegraphics[width=0.98\textwidth]{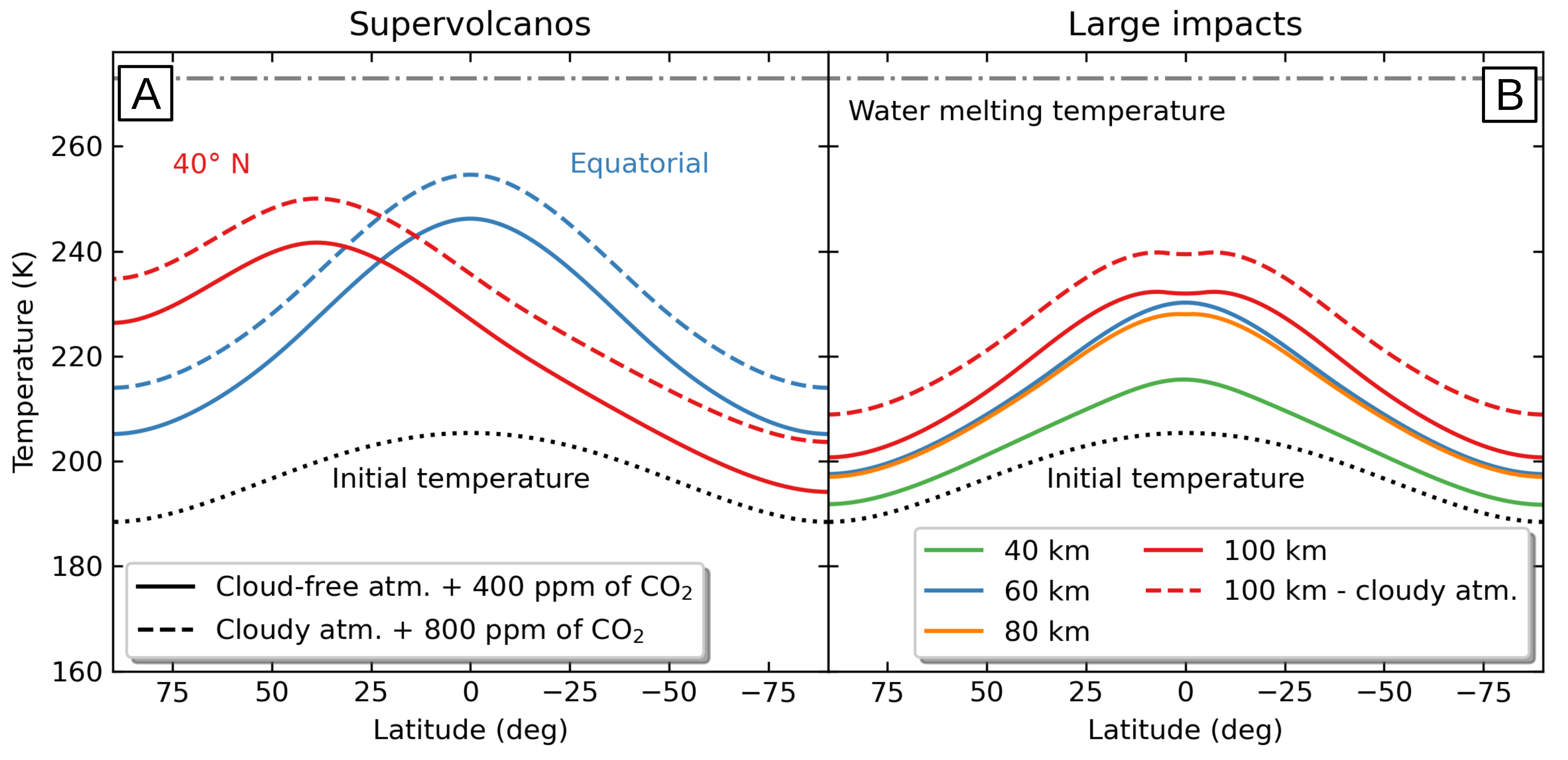}
    \caption{Latitudinal temperature profiles after (A) Toba-like eruptions, both at the equator (blue lines) and at $40^\circ \;\mathrm{N}$ (red lines), and with both the cloud-free atmosphere with $400\;\mathrm{ppm}$ \ce{CO2} (solid lines) and an atmosphere including cloud radiative effects and $800\;\mathrm{ppm}$ \ce{CO2}, and (B) impacts with impact diameters of $40 \;\mathrm{km}$ (green), $60 \;\mathrm{km}$ (blue), $80 \;\mathrm{km}$ (orange), and $100 \;\mathrm{km}$ (red), all striking at the equator and under cloud-free atmosphere with $400\;\mathrm{ppm}$ \ce{CO2} (solid lines). The red dashed line in panel B corresponds to a $100 \;\mathrm{km}$ impactor under a cloudy atmosphere, assuming $400\;\mathrm{ppm}$ of \ce{CO2}. The initial temperature profile (i.e., before climate forcing stochastic events) is shown for both types of events (dotted lines).
    } 
    \label{fig:converged}
\end{figure}

The equilibrium spatial temperature profiles (i.e., spatially resolved profiles for the last times shown in Figure~\ref{fig:temp_evol}A-H) can provide further insight into the climate forcing effects of our stochastic events (Figure~\ref{fig:converged}). 
Without any stochastic forcing, we observe the usual latitudinal distribution of surface temperatures, with a maximum of $\sim205 \;\mathrm{K}$ at the equator and a minimum of $\sim 190 \;\mathrm{K}$ at the poles.

For a Toba-like eruption, we find that deposition of the volcanic ash leads to greatest temperature rises near the deposition site itself (Figure~\ref{fig:converged}A). Ash deposition affects the surface albedo on a limited area, locally increasing the absorbed flux. 
For this reason, considering the eruption at $40^\circ \,\mathrm{N}$ (Figure~\ref{fig:converged}A, red line) centers the peak of the temperature profile to this location, and lower the maximum temperature of about $5 \;\mathrm{K}$ due to a weaker incoming flux at this latitude.
Additionally, for both cases, heat diffusion (atmospheric circulation) allows a global increase of temperature.
By accounting for the radiative effect of the clouds and by assuming $800 \;\mathrm{ppm}$ of \ce{CO2}, the extra warming is around $10 \;\mathrm{K}$, independent of latitude and of the eruption location.
We note that the results shown here do not account for the ash injected into the stratosphere, which reduces the bond albedo of the planet and thus induces global cooling \cite<e.g.,>{abbot_mudball_2010}. We discuss such processes, as well as others that we have chosen to neglect in the presented results, and their effect on our conclusions in Section~\ref{sec:discussion}. 

For impact events, the converged temperature profiles are warmer than the non-perturbed simulation (Figure~\ref{fig:converged}B), due to ejecta affecting the surface albedo. The rise in temperature is a function of the area covered by ejecta: the greater the area covered, the more the albedo of the surface is affected. However, the maximum temperatures for all impact scenarios are colder than those found in scenarios forced by the supervolcanic eruptions. The injection of heat that is provided by the impactors is a relatively minor effect in comparison to the change in surface albedo of the ice that is brought about by the ejecta curtain. Evidence for this can also be found in the temporal evolution of temperatures after the impacts (Figure~\ref{fig:temp_evol}E-H). In the four cases, the large radiative imbalance at the impact site dissipates heat rapidly relative to the timescale of global temperature changes. Therefore, the colder converged temperatures of our impacts are due to the diminished ejecta coverage that they produce in comparison to the coverage of the supervolcanic ash (see Supplementary Figure S3). Interestingly, for $80 \;\mathrm{km}$ and $100 \;\mathrm{km}$ impactors, we find a cold patch in the converged temperature profile at the impact site. We determine this to be due to the transient water belt that forms after the impact: when the planet cools, this liquid water freezes at the impact site as clean ice with greater albedo than the ejecta-covered ice around the crater. 

\section{Discussion} \label{sec:discussion}
\subsection{Limitations of the climate model}
\ce{H2O} and \ce{CO2} are efficient greenhouse gases and should facilitate the warming and deglaciation of the ice sheet. However, we find that variation of the gas partial pressures, within sensible bounds for the climate of a Neoproterozoic Snowball Earth, cannot push our stochastic events into regimes where they cause global deglaciation (see Supplementary Text S5). 

There are a variety of effects that are not included in the EBM that could be considered as important. We neglect the formation of stratospheric aerosols, which would act to cool the planet \cite{macdonald_initiation_2017}. In order to minimize the degrees of freedom in the model, we do not account for ice thickness or the latent heat released by ice melting, which could be important in accounting for the time-scales over which the water belt induced by the $100$-$\mathrm{km}$ impactor freezes. In much the same way, oceanic circulation will affect the creation of ice in the water belt \cite<e.g.,>{yang_transition_2019, brunetti_co-existing_2019}. However, a commonality between these effects is that their inclusion would serve only to increase the challenge in producing global deglaciation, which our models already suggest is not achievable given the scales of events considered. We thus suggest that our simplified EBM is able to point out possible directions for further works using more complex models.

\subsection{Assumptions on impacts and large eruptions}
Our simulations suggest that impacts larger than $100 \;\mathrm{km}$ are necessary, albeit not sufficient, to initiate deglaciation on the planet. However, dynamical models indicate that, outside of the early Solar System, such impacts are expected to occur less than once every billion years \cite{koeberl2019asteroid}. Moreover, geologic evidence for such event should be present, since a large projectile should have globally spread an ejecta layer analogous to the Chicxulub impact event \cite{Alvarez1980}. So far, no impact signature caused by a $\sim 100$-$\mathrm{km}$ impactor has been found on Earth \cite{gyollai2014lack,PEUCKEREHRENBRINK2016}; as per \citeA{allen2022}, the largest impact crater found is caused by a $20-25 \,\mathrm{km}$ impactor.

The results we present are based on a single impact scenario, comprising of a fixed stratigraphy and an asteroidal projectile hitting the surface at $17 \;\mathrm{km \,s^{-1}}$. We chose an ice thickness of $5 \;\mathrm{km}$ following \citeA{Erickson2020} as a reasonable upper limit for the actual size of glaciers covering the planet. In reality, the ice thickness will depend on the location, as suggested by the geologic record \cite<e.g.,>{McMechan2000} and climate modeling \cite{Hyde2000}, ranging from meters to $5 \;\mathrm{km}$. Variation in ice thickness will produce a mechanically different response to impact. However, given the scale of impactor that our models find would be necessary to cause global deglaciation (i.e., $d > 100 \;\mathrm{km}$), the ice thickness is relatively shallow in comparison, and we thus expect that the consequences of this effect would be minor. Further, we did not consider impactors of alternative compositions, such as icy bodies, in our models. While short-period comets \cite<impact velocity of $30 \;\mathrm{km \,s^{-1}}$, >{CHYBA1991} have kinetic energies similar to that of the asteroidal impacts we consider, long-period comets \cite<$50 \;\mathrm{km \,s^{-1}}$, >{CHYBA1991} might change the impact mechanics, but are less likely to hit the Earth \cite{weissman_2006}. Such events would leave behind a distinct geochemical signal from a rocky impactor analogue that, while not yet discovered, should similarly perhaps not be excluded for now. Further modeling of such stochastic events could be entertained by future works, although will likely find challenges of scale similar to our work (i.e., requiring extremely unlikely comet sizes to cause deglaciation). Finally, we assumed zero porosity as a worst case scenario: for a given size, non-porous material ensures the maximum kinetic energy available to melt ice and vaporize target rocks. Similar to the climate choices made, therefore, relaxing this assumption only serves to make deglaciation more challenging.

In modeling volcanic eruptions, results are sensitive to the assumed dust distribution and the absence of lava warming and water vapor. Based on Icelandic dust and ash \cite{Dragosics2016}, a thickness of $15 \;\mathrm{mm}$ is considered as the threshold for ice to melt due to the decrease in albedo. A thicker layer would thermally insulate the ice below, preventing melting. Increasing such a threshold would expand the dust-covered area subject to melting, favoring deglaciation. Furthermore, the local temperature increase due to erupting lava is disregarded, since its effect is negligible when averaged longitudinally. Similarly, we do not consider a long-term (i.e., multi-year) eruption, which would force a higher temperature around the volcano for a longer time. Additionally, as seen for the Hunga Tonga-Hunga Ha'apai eruption, a substantial amount (up to $10^{11} \;\mathrm{kg}$) of water vapor can be injected into the stratosphere where it can have multi-year long radiative effects \cite<net warming of the surface, >{Millan2022}. Finally, aerosols ejected up to the stratosphere could increase snowfall, thus covering low albedo dusty ice.

\subsection{Possible shortening of snowball events}
In this work, we consider a low \ce{CO2} pressure as the initial climate state. However, during a snowball event the carbon cycle is altered due to the ice shield preventing \ce{CO2} (partial) dissolution into the ocean and restricted silicate weathering on land. On modern Earth, \ce{CO2} outgassing is about 47~bar/Gyr \cite{catling_atmospheric_2017} through a combination of continental and submarine volcanism. Although there is no consensus so far on the exact volcanic outgassing flux for snowball Earth, multiple studies suggest that the value was much lower than present day. 
\citeA{fischer_agu_2020} estimate the arc production around 5.7~bar/Gyr while \citeA{dutkiewicz_duration_2024} argue that the mid-ocean ridge outgassing was even lower. 
We found that an extra radiative forcing of 50~W/m$^2$ (in addition to 14~W/m$^2$ of forcing induced by the clouds) is required, coupled with the effect of a 100~km impactor, to deglaciate the planet. That is corresponding approximately to 0.08~bar of \ce{CO2} according to \citeA{wolf_evaluating_2018}. In comparison, without any stochastic event the required forcing to deglaciate is equal to 90~W/m$^2$ (in addition to 14~W/m$^2$ of forcing induced by the clouds).
Using modern day Earth outgassing from \citeA{fischer_agu_2020} as boundary values, it takes between 1.7~Myr and 14~Myr to reach a \ce{CO2} pressure for which a 100-km impactor could induce a deglaciation, in our framework. An intermediate outgassing value of 25~bar/Gyr gives approximately 3.2~Myr to reach the required \ce{CO2} pressure.

These estimates should be considered as lower limits because we neglect the potential sinks of \ce{CO2} existing on snowball Earth, as well as cooling processes of impacts (e.g., increase of the bond albedo due to aerosols). However, this timescale is shorter than the duration time of the Marinoan and Sturtian glaciations. Although a large impact during this period seems highly unlikely, the possibility of deglaciating an extra-solar planet in this way cannot be ruled out.

\section{Conclusions} \label{sec:conclusions}
\noindent
Earth's periods of global glaciation are traditionally conceived to have ended through a gradual build-up greenhouse gases in the atmosphere (e.g., \ce{CO2}). However, the shortest of these so-called Snowball Earth states were geologically short and stochastic events such as impacts and supervolcanic eruptions has been suggested as plausible deglaciation precesses. Here, we present modeling on the response of Snowball Earth to such events under differing scenarios. We use impact simulations and ash spreading models to estimate the initial effect of each stochastic event, and follow these up by evolving Earth's surface environment through time using a 1-D EBM. 

In our modeling framework, surface albedo of the planet is the dominant driving force of long-term temperature change in the aftermath of stochastic events. For impacts, this takes the form of the ejecta curtain stemming from the impact site. For eruptions, volcanic ash is transported over a wide range of latitude and longitudes. We thus observe that eruptions produce a greater change in Earth's surface albedo and hence a greater temperature response for typical scale events. However, even an impactor radius of $100 \;\mathrm{km}$ is not sufficient to induce deglaciation of Snowball Earth. The warming at the impact site melts ice locally in the short-term, but due to a large radiative local imbalance, ice reforms in less than a year, returning the planet to the snowball state. Even if larger impactors could induce deglaciation, they are dynamically unlikely to have occurred on Earth at the time of its snowball episodes. Similarly, no recorded supervolcanic eruption could have single-handedly deglaciated Earth. The surface temperature increase due to dust deposition from the largest recorded event, Toba, does not reach the melting temperature of ice anywhere on the planet. 

Our model thus places constraints on the magnitude of stochastic events that are incapable of triggering Snowball Earth termination. However, these constraints should be refined with more comprehensive modeling. For impacts, 3-D modeling can better resolve the ejecta distribution through accounting for oblique impacts. Impact simulations should also be able to account for the ice erosion effect due to the tsunami created upon impact, which would increase the low-albedo area. For the climate response, 3-D GCM modeling is required to include longitudinal and vertical resolutions, as well as to provide better representations of clouds, rain/snow, atmospheric dynamics, ocean dynamics, and distribution of aerosols ejected by the impactor \cite<e.g.,>{abbot_jormungand_2011,ashkenazy_dynamics_2013,turbet_environmental_2020} or released by the volcano \cite{mcgraw_severe_2024}. For instance, even 2-D models could show more complex transitions between two steady states if specific eigenmodes are triggered by an impact happening at the right location \cite{mulder_snowball_2021,bastiaansen_fragmented_2022}.

Lastly, in this work, we treated impacts and supervolcanoes as independent and instantaneous events all starting from the same initial state. Future analyses should investigate some of the prolonged effects associated with our stochastic events (e.g., multi-year lava eruptions). Additionally, the combined effect of impacts and eruptions should be considered, with impact-induced volcanism having been suggested as a potential supplement for the K-T extinction \cite<e.g.,>{Renne2015}, although still debated \cite<e.g.,>{Bhandari1995}. Multiple such events would likely stack their temperature responses, although unlikely to be in a linear manner, and could thus instigate a deglaciation of Snowball Earth.

\section*{Open Research}
\noindent
At present, the iSALE code \cite{Collins2016} is not fully open-source. It is distributed via a private GitHub repository on a case-by-case basis to academic users in the impact community, strictly for non-commercial use. 
\\ \\
The temperature profile after impact, used to perform climate simulations can be found from \citeA{chaverot_resilience_2024}.

\acknowledgments 
The project was made possible by the 2022 Cooperative Institute for Dynamic Earth Research (CIDER) workshop in UC Berkeley, funded by the NSF. This work has been carried out within the framework of the NCCR PlanetS supported by the Swiss National Science Foundation under grants 51NF40\_182901 and 51NF40\_205606. GC acknowledges the financial support of the SNSF (grant number: 200021\_197176, 200020\_215760 and P500PT\_217840). AZ acknowledges the financial support of Stanford University. JI was supported by the UK Science and Technology Facilities Council (STFC) grant number ST/T505985/1. RG acknowledges funding by the National Aeronautics and Space Administration (NASA) through a contract with ORAU.
\\
\\
We gratefully acknowledge the developers of iSALE-2D, including Gareth Collins, Kai Wünnemann, Dirk Elbeshausen, Tom Davison, Boris Ivanov and Jay Melosh.
\\
\\
We thank the two anonymous reviewers for the time they spent providing useful comments that helped to improve this manuscript.
Authors gratefully acknowledge R. T. Pierrhumbert for useful discussions during the 2022 CIDER workshop. 
\\ 
\\
Authors declare that they have no competing interests.

\noindent \textbf{Author contributions}
\begin{itemize}
    \item[] Conceptualization: GC, BF
    \item[] Methodology: AZ, GC, SB, XD, BF, JI, AJ, TM
    \item[] Investigation: GC
    \item[] Project administration: AZ, GC
    \item[] Writing – original draft: AZ, GC
    \item[] Writing – review \& editing: AZ, GC, XD, BF, SB, JI, TM, RG
    \item[] Data Curation: AZ, GC, JI
\end{itemize}


\bibliography{biblio}





\section*{Supplementary material}

**Contents of this file**

1. Texts S1 to S6

2. Figures S1 to S3

3. Tables S1 to S3

**Introduction**

The following Supplementary Information provides first an analytical calculation of the order of magnitude of energy required to deglaciate Snowball Earth (Text S1). details about numerical setups of both the impacts simulations (Text S2, Table S2) and EBM simulations (Text S4, Table S3). Also, we provide additional discussion on the approximations made for impact simulations (Text S3), on the sensitivity texts done on the EBM setup (Text S5), as well as a discussion concerning implications of this work for exoplanets (Text S6).

**Text S1: Analytical estimation of energy required for deglaciation**

It is possible to estimate the order of magnitude of energy required to deglaciate the snowball Earth, using a simple analytical calculation. To do so, we first estimate equilibrium climates with a 0-D energy balance equation:

$$(1 - A) F = a + bT_s \tag{1}$$

where $A$ is global mean albedo (reflectivity), $F$ is global mean incoming solar flux, $T_s$ is global mean surface temperature, and $a$ and $b$ are constants. The left handside represents absorbed solar insolation; the right handside represents emitted infrared radiation.

―――――

Following Koll and Cronin (2018) and Wagner and Eisenman (2015), the parameters are chosen as $A = \begin{cases} A_o = 0.3, & T > T_m \\ A_i = 0.6, & T < T_m \end{cases}$, where $T_m = 273$ K is the melting point of ice, $a = -339.647$ W/m$^2$, and $b = 2.218$ W/m$^2$/K. Thus, 3 possible scenarios exist based on the choice of incoming solar flux $F$. When $F$ is too small or too large, only one solution for climate can be found. When $F$ is moderate, three solutions can be found. The snowball bifurcation is represented by the warmest and coldest solutions (the middle solution $T_m$ is unstable). We find the range of $F$ that allows the snowball bifurcation is $381.2$ W/m$^2 < F < 667.1$ W/m$^2$.

Then we estimate the energy required for snowball deglaciation by picking up a typical solar forcing within the bifurcation range:

$$F = \sqrt{F_1 F_2} = (a + bT_m)\sqrt{\frac{1}{(1 - A_i)(1 - A_o)}} = 502 \text{ W/m}^2 \qquad (2)$$

The two stable solution can be solved analytically:

$$T_{snowball} = \frac{(1 - A_i)F - a}{b} = \frac{a}{b}\left(\sqrt{\frac{1 - A_i}{1 - A_o}} - 1\right) + T_m\sqrt{\frac{1 - A_i}{1 - A_o}} = 243.7 \text{ K} \qquad (3)$$

$$T_{warm} = \frac{(1 - A_o)F - a}{b} = \frac{a}{b}\left(\sqrt{\frac{1 - A_o}{1 - A_i}} - 1\right) + T_m\sqrt{\frac{1 - A_o}{1 - A_i}} = 311.7 \text{ K} \qquad (4)$$

To transition from the snowball state $T_{snowball}$ to $T_{warm}$, we only need to heat the planet over the unstable point $T_m$. Once $T > T_m$, the planet will converge to $T_{warm}$ within finite time. The energy required for snowball deglaciation is:

$$Q_d = C\left(T_m - T_{snowball}\right) \qquad (5)$$

where $C$ is the heat capacity of climate system. Assuming the heat capacity mainly comes from the mixed-layer ocean, $C$ can be estimated as:

$$C = c_s \alpha 4\pi r^2 h_{mix} \rho = 6.0 \times 10^{22} \text{ J/K} \tag{6}$$

where $c_s = 4184$ J/kg/K is the specific heat of water, $\alpha = 0.7$ is surface fraction of ocean on Earth, $r$ is Earth's radius, $h_{mix} = 40$ m is the global-mean thickness of mixed layer, and $\rho = 10^3$ km/m$^3$ is density of water.

Based on the estimation of $C$, the minimum energy for snowball deglaciation $Q_d$ is

$$Q_d = C\left(T_m - T_{snowball}\right) = 1.76 \times 10^{24} \text{ J} \tag{7}$$

Additional energy is required if including the latent heat flux (phase change) to melt the ice. Following Pierrehumbert, Abbot, Voigt, and Koll (2011), the energy to melt the ice $Q_{ice}$ is $1.7 \times 10^{26}$ J assuming a global ice layer controlled by geothermal heat flux of 0.05 W/m$^2$ and a vertical temperature difference of 30 K within the ice. $Q_{ice}$ is two orders of magnitude larger than $Q_d$.

Then we connect $Q$ with the size of impact $D$. We use Online crater scaling tools (https://impact.ese.ic.ac.uk/ImpactEarth/ImpactEffects/ and https://www.eaps.purdue.edu/impactcrater/crater_c.html). The input parameters are projectile density equal to $10^3$ kg/m$^3$, impact velocity of 17 km/s, impact angle of 45°, target density equal to $10^3$ km/m$^3$ and Earth gravity. The results are listed in Table S1. The 40-km and 100-km impacts are barely enough to overcome $Q_d$ but not $Q_{ice}$. A 400-km impact overcomes both $Q_d$ and $Q_{ice}$. Thus, impact size > 100 km ($Q > 10^{26}$ J) is preferred to generate a snowball-deglaciating impact.

**Text S2: Impact simulation setup**

The iSALE shock physics code (Wünnemann et al., 2008) is an extension of the SALE hydrocode (Amsden et al., 1980). Modifications to the original SALE code include an elasto-plastic constitutive model, fragmentation models, various equations of state, and multiple materials (Melosh et al., 1992; Ivanov et al., 1997). More recent improvements include a modified strength model (Collins et al., 2004), a porosity-compaction model (Wünnemann et al., 2006; Collins et al., 2011), and a dilatancy model (Collins, 2014). The thermodynamic behavior of each material is described by equations of state generated by the ANalytic Equation Of State program (ANEOS; Thompson & Lauson, 1974). The detailed properties of each material and the relevant equations of state are detailed in Table S1.

Our impactors are sufficiently large that the curvature of the Earth is important. However, for the information we wish to extract from the simulations, it is still appropriate for them to remain 2-D, which assists in the trade-off between resolution and computational resources.

**Text S3: Discussion on impact modeling approximations**

Because of the strong effect of albedo on the climate evolution, our conclusions are sensitive to the assumptions made in estimating the dust distribution, namely assuming vertical impacts and no atmosphere present in the impact model. With a 2-D hydrocode, we are unable to reproduce oblique collisions, which affects (1) the amount of vaporized material, since shock waves would propagate in a larger volume of ice/crust in oblique impacts compared to vertical ones with the same kinetic energy, and (2) the shape of ejecta deposits, with oblique impacts preferentially distributing in the downrange direc-

tion (Collins et al., 2020). Three-dimensional simulations are needed to better understand whether oblique impacts can cause a larger dust coverage. However, we again suggest that conclusions surrounding deglaciation would remain the same well within order of magnitude. In terms of not including an atmosphere in our impact simulations, the main effect is that ejecta follows ballistic trajectories that are unperturbed by atmospheric circulation. Rock vapor is thus deposited on the surface without residing in the atmosphere. Consequently, the extension of dust we estimate is a lower limit, since we neglect spreading at further distances due to atmospheric transport (e.g., Kring, 2007). This would thus serve to decrease the planetary albedo, and should not affect our conclusions.

**Text S4: EBM parameterization**

We investigate the climate of the model planets using a 1-D time-dependent EBM introduced in Williams, Kasting, and Frakes (1998), Spiegel, Menou, and Scharf (2008) and further explored by Dressing, Spiegel, Scharf, Menou, and Raymond (2010). The equation governing the temporal evolution of the surface temperature $T$ is given by:

$$C\frac{\partial T[x,t]}{\partial t} - \frac{\partial}{\partial x}\left(D(1-x^2)\frac{\partial T[x,t]}{\partial x}\right) + I[T,t] = S(1 - A[T,t]), \quad (8)$$

where $x = \sin\lambda$ gives the location of the grid point of latitude $\lambda$, $t$ is time, $C$ is the effective heat capacity of the atmosphere over this grid point, $D = 0.5394$ W m$^{-2}$ K$^{-1}$ is the diffusion coefficient which determines the efficiency of the latitudinal redistribution of heat, $I(T,t)$ is the temperature-dependent infrared emission, $S$ is the incoming stellar radiation and $A(T,t)$ is the temperature-dependent albedo of the grid point. The net radiative energy flux at a grid point is determined by the relationship between the irradiation

CHAVEROT ET AL.: SNOWBALL EARTH RESILIENCE TO STOCHASTIC EVENTS    X - 7

arriving on the planet ($S$ in Eq. (8)) and the energy lost via infrared emission ($I$ in Eq. (8)).

Following Spiegel et al. (2008) and Dressing et al. (2010), we compute the absorbed stellar radiation (ASR) as a function of the surface albedo $A$ and do not account for the atmospheric dependency. The albedo varies from 0.28 for water (accounting for Earth's continents) to 0.77 for ice following Eq. (9):

$$A = 0.525 - 0.245 \tanh\left(\frac{T - 268}{5}\right) \qquad (9)$$

We adapt Eq. (9) to compare our simulations with Pierrehumbert et al. (2011), who assumed 0.5 for the albedo of the ice. The albedo of "liquid water" is fixed to 0.28 to account for continents (Dressing et al., 2010). The equation giving the insolation as a function of the latitude with obliquity is taken from Berger (1978) assuming no eccentricity. The heat capacity of the ground depends on the surface type of the grid point and varies for land, ocean, ice and cold ice, following Spiegel et al. (2008) and Dressing et al. (2010).

For the thermal emission, we consider two different parameterization made using the 1D radiative-convective model presented in Chaverot, Turbet, Bolmont, and Leconte (2022), representing the atmosphere of the Earth with 1 bar of nitrogen and 400 ppm or 800 ppm of $CO_2$. In both cases, the water vapor pressure is a function of the temperature following the pseudo-adiabatic assumption (Kasting, 1988). This original parameterization allow us to account for $CO_2$ released by volcanic eruptions. The relative humidity (RH) is constant during the simulation and can be equal to 0.45 or 1. As shown by Leconte,

Forget, Charnay, Wordsworth, and Pottier (2013), RH=0.45 is able to reproduce the infrared emission of a temperate Earth from a Global Climate Model. Higher RH values are used to represent the extra moisture accumulated in the atmosphere after a large eruption. This original parameterization of the thermal emission provides similar results to those obtained using the analytical parameterization from Dressing et al. (2010). The temperature profiles of the temperate and snowball regimes are shown in Fig. S2 (green and blue lines, respectively). Additionally, a stable water-belt configuration is shown for a diffusion coefficient $D = 0.25$ W m$^{-2}$ K$^{-1}$ (orange line). These different equilibrium states are obtained for a same setup by assuming different initial temperatures, highlighting climate multistability.

By definition, the EBM is longitudinally-averaged which implies that modifying one grid cell affects all corresponding longitudes. As the considered stochastic events are limited to a particular region in space, we add a correction factor so that not all longitudes are impacted. More specifically, during the initialization of the simulation, we average the values of the albedo and of the temperature after the event with the non-impacted part of the planet at every latitude, accounting for the area of this event. By doing so, the initial maximum temperature of the EBM is much lower than the maximum temperature after the impact obtained from iSALE.

**Text S5: Sensitivity tests**

We perform several sensitivity tests in order to benchmark the original parameterization we introduced in the EBM for the thermal emission and to quantify the intrinsic effect of

specific parameters. An exhaustive overview of these tests is presented in Table S3 and an overview of the most important ones is shown in Fig. S1.

First, the temperature profile we obtain using our parameterization of the thermal emission with 400 ppm of $CO_2$ is consistent with that of Dressing et al. (2010). Second, we tested the impact of several parameters of the EBM. The value of the diffusion factor ($D$ in Eq. (8)) we used is taken from Dressing et al. (2010), but as mentioned in Pierrehumbert et al. (2011), cold climates have a reduced heat transport efficiency. Therefore, we tested $D = 0.25$ W m$^{-2}$ K$^{-1}$ as per Pierrehumbert et al. (2011) (see orange curve in Fig. S1). The modification of the heat transport increases the temperature contrast between the equator and poles by a few Kelvin. We also added a cloud forcing equal to 14 W m$^{-2}$ to account for their warming effect, according to Abbot (2014). This increases the global temperature by 10 K (difference between blue and red curves in Fig. S1).

Figure S1 also shows that by accounting for 800 ppm of $CO_2$ instead of 400 ppm, the temperature increases by less than 1 K. Likewise, increasing RH up to 1 does not significantly change the temperature from the point of view of deglaciation point. This is coherent regarding the low water vapor content of a cold atmosphere. Only tests inducing a significant variation of the temperature profile are shown on Fig. S1.

Due to our longitudinal correction, we overestimate the albedo by less than 1% depending on the albedo difference and on the temperature gradient along the longitude.

**Text S6: Implications for Earth-like exoplanets**

While the modeling in this study is predicated on Earth-like conditions, our conclusions may yet be useful for considerations of Earth-like exoplanets experiencing similar

phenomena to Snowball Earth. For example, the question of planetary bi-stability within exoplanet demographics may be of particular relevance: if deglaciation through stochastic events is commonplace, the non-glacial planetary state will be heavily favored.

There are a variety of model components that may vary between worlds, such as stratigraphy, volcanic dust spreading through atmospheric circulation, atmospheric chemical composition and thermal structure, geologic record for impact and volcanic events, tectonic regime, and impactor size-frequency distributions. For instance, atmospheric dynamics and structure on planets with thicker or more turbulent atmospheres may produce ash spreading far beyond that found for Toba, enabling greater albedo change. Similarly, the thermo-mechanical response of basalt is different from that of granite (which we assumed to be the material underneath the ice layer), meaning that alternate crustal compositions and structures may produce wider dust spreading and/or surface temperature profiles. Finally, and perhaps most importantly, limiting factors on the scales of events for Earth do not necessarily apply to other planets. Supervolanic eruptions of greater magnitude than Toba may be more common, as may be projectiles of 100 km diameter or more.

Therefore, while we suggest that stochastic events are unlikely to have caused deglaciation of Snowball Earth, we do not rule out such events may be able to cause deglaciation on worlds where the parameters permit it.

**Table S1.** Connection between impact energy $Q$ and impact size $D$

| D (km) | Q (J) |
|---|---|
| 40 | $4.84 \times 10^{24}$ |
| 100 | $7.57 \times 10^{25}$ |
| 400 | $4.84 \times 10^{27}$ |

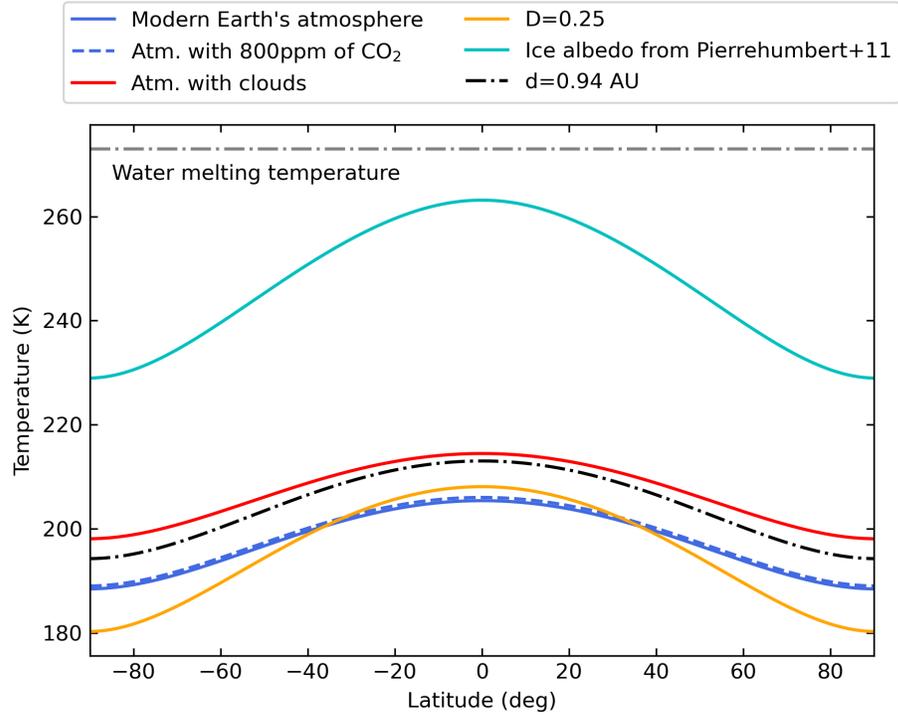

**Figure S1.** Latitudinal temperature without stochastic events for different parameterizations. Solid curves are for the reference thermal emission corresponding to the Earth's atmosphere including 400 ppm of $CO_2$. The dashed curve corresponds to simulations with 800 ppm of $CO_2$. Red, orange and cyan curves are for simulations with cloud forcing, with a diffusion factor $D = 0.25$ W m$^{-2}$ K$^{-1}$ and an ice albedo from Pierrehumbert et al. (2011) respectively. The black dash-dotted curve is for the Earth located at 0.94 AU from the Sun.

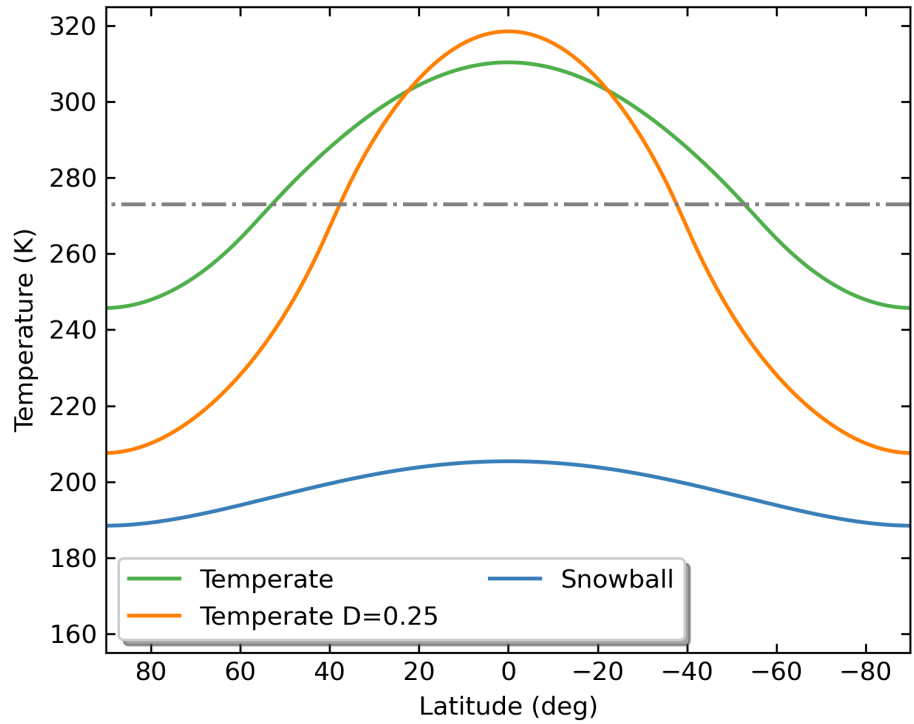

**Figure S2.** Latitudinal temperature of the snowball (blue line) and temperate (green line) steady states. The orange line corresponds to a water-belt configuration induced by a lower diffusion coefficient (here $D = 0.25$ W m$^{-2}$ K$^{-1}$).

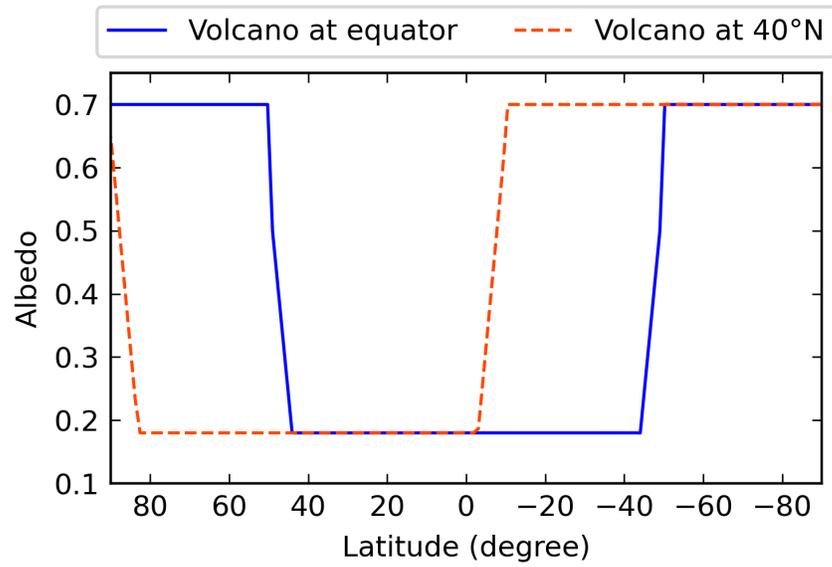

**Figure S3.** Latitudinal albedo parameterized for Toba-like supervolcanic eruptions with the eruption center at the equator (solid curve) and at 40°N (dashed curve). The albedo values of ice (0.7), dusty ice (0.18-0.7) and dust (0.18) are based on Le Hir et al. (2010) and Warren (1982).

**Table S2.** Material parameters used within simulations for the impactors and targets. All parameters used are standard to iSALE or M-ANEOS, and details can be found in the respective manuals (iSALE: `https://github.com/isale-code/isale-wiki`; M-ANEOS: `https://github.com/isale-code/M-ANEOS`).

| Parameter | Dunite | Graphite | Calcite | $H_2O$ Ice |
|---|---|---|---|---|
| EOSNAME | dunite_ | granite | calcite | ice___ |
| No. density nodes | 240 | 240 | 240 | 240 |
| No. thermal nodes | 120 | 120 | 240 | 240 |
| Low density weakening | POLY | POLY | POLY | POLY |
| Poisson ratio | $2.5 \times 10^{-1}$ | $3.0 \times 10^{-1}$ | $3.0 \times 10^{-1}$ | $3.3 \times 10^{-1}$ |
| Strength model | ROCK | ROCK | ROCK | ICE |
| YINT0 | $1.0 \times 10^7$ | $1.0 \times 10^7$ | $5.0 \times 10^6$ | $1.0 \times 10^7$ |
| FRICINT | 1.2 | 2.0 | 1.0 | 2.0 |
| YLIMINT | $3.5 \times 10^9$ | $2.5 \times 10^9$ | $5.0 \times 10^8$ | $1.1 \times 10^8$ |
| YDAM0 | $1.0 \times 10^4$ | $1.0 \times 10^4$ | $1.0 \times 10^4$ | $5.0 \times 10^5$ |
| FRICDAM | $6.0 \times 10^{-1}$ | $6.0 \times 10^{-1}$ | $4.0 \times 10^{-1}$ | $6.0 \times 10^{-1}$ |
| YLIMDAM | $3.5 \times 10^9$ | $2.5 \times 10^9$ | $5.0 \times 10^8$ | $1.1 \times 10^8$ |
| Damage model | IVANOV | | | |
| IVANOV_A | $1.0 \times 10^{-4}$ | | | |
| IVANOV_B | $1.0 \times 10^{-11}$ | | | |
| IVANOV_C | $3.0 \times 10^8$ | | | |
| Thermal softening | OHNAKA | OHNAKA | OHNAKA | OHNAKA |
| TMELT0 | $1.373 \times 10^3$ | $1.673 \times 10^3$ | $1.500 \times 10^3$ | $0.272 \times 10^3$ |
| TFRAC | 1.2 | 1.2 | 1.2 | 1.2 |
| ASIMON | $1.52 \times 10^9$ | $6.0 \times 10^9$ | $6.0 \times 10^9$ | $6.0 \times 10^9$ |
| CSIMON | 4.05 | 3.00 | 3.00 | 3.00 |

**Table S3.** Overview of the EBM setups

| Simulation | $CO_2$ pressure | RH | Extra param. |
|---|---|---|---|
| No stochastic event | 400 ppm | 0.45 | |
| No stochastic event | 400 ppm | 0.45 | Ice albedo $= 0.5$ (Pierrehumbert et al., 2011) |
| No stochastic event | 400 ppm | 0.45 | $d = 0.94$ AU |
| No stochastic event | 400 ppm | 0.45 | $D = 0.25$ W m$^{-2}$ K$^{-1}$ |
| No stochastic event | 400 ppm | 0.45 | obliquity $\phi = 23°$ |
| No stochastic event | 400 ppm | 0.45 | cloud forcing $= 14$ W m$^{-2}$ (Abbot, 2014) |
| No stochastic event | 400 ppm | 1 | |
| No stochastic event | 800 ppm | 0.45 | |
| Toba at the equator | 400 ppm | 0.45 | |
| Toba at the equator | 800 ppm | 0.45 | |
| Toba at the equator | 400 ppm | 0.45 | cloud forcing $= 14$ W m$^{-2}$ (Abbot, 2014) |
| Toba at the equator | 800 ppm | 0.45 | cloud forcing $= 14$ W m$^{-2}$ (Abbot, 2014) |
| Toba at 40° N | 400 ppm | 0.45 | |
| Toba at 40° N | 800 ppm | 0.45 | |
| Toba at 40° N | 400 ppm | 1 | |
| 40 km impactor | 400 ppm | 0.45 | |
| 60 km impactor | 400 ppm | 0.45 | |
| 80 km impactor | 400 ppm | 0.45 | |
| 100 km impactor | 400 ppm | 0.45 | |
| 100 km impactor | 400 ppm | 1 | |
| 100 km impactor | 800 ppm | 0.45 | |
| 100 km impactor at 40° N | 400 ppm | 0.45 | |

July 12, 2024, 5:54pm

**References**

Abbot, D. S. (2014). Resolved Snowball Earth Clouds. *Journal of Climate*, *27*(12), 4391–4402. doi: https://doi.org/10.1175/JCLI-D-13-00738.1

Amsden, A. A., Ruppel, H. M., & Hirt, C. W. (1980). *SALE: a simplified ALE computer program for fluid flow at all speeds* (Tech. Rep.). Los Alamos Scientific Laboratories. doi: https://doi.org/10.2172/5176006

Berger, A. (1978). Long-Term Variations of Daily Insolation and Quaternary Climatic Changes. *Journal of the Atmospheric Sciences*, *35*(12), 2362–2367. doi: https://doi.org/10.1175/1520-0469(1978)035⟨2362:LTVODI⟩2.0.CO;2

Chaverot, G., Turbet, M., Bolmont, E., & Leconte, J. (2022). How does the background atmosphere affect the onset of the runaway greenhouse? *Astronomy & Astrophysics*, *658*, A40. doi: https://doi.org/10.1051/0004-6361/202142286

Collins, G. S. (2014). Numerical simulations of impact crater formation with dilatancy. *Journal of Geophysical Research: Planets*, *119*(12), 2600–2619. doi: https://doi.org/10.1002/2014JE004708

Collins, G. S., Melosh, H. J., & Ivanov, B. A. (2004). Modeling damage and deformation in impact simulations. *Meteoritics & Planetary Science*, *39*(2), 217–231. doi: https://doi.org/10.1111/j.1945-5100.2004.tb00337.x

Collins, G. S., Melosh, H. J., & Wünnemann, K. (2011). Improvements to the $\varepsilon - \alpha$ porous compaction model for simulating impacts into high-porosity solar system objects. *International Journal of Impact Engineering*, *38*(6), 434–439. doi: https://doi.org/10.1016/j.ijimpeng.2010.10.013

CHAVEROT ET AL.: SNOWBALL EARTH RESILIENCE TO STOCHASTIC EVENTS          X - 17Collins, G. S., Patel, N., Davison, T. M., Rae, A. S. P., Morgan, J. V., Gulick, S. P. S., ... Scientists, T.-P. (2020). A steeply-inclined trajectory for the Chicxulub impact. *Nature Communications*, *11*(1), 1480. doi: https://doi.org/10.1038/s41467-020-15269-x

Dressing, C. D., Spiegel, D. S., Scharf, C. A., Menou, K., & Raymond, S. N. (2010). Habitable Climates: The Influence of Eccentricity. *The Astrophysical Journal*, *721*, 1295–1307. doi: 10.1088/0004-637X/721/2/1295

Ivanov, B. A., Deniem, D., & Neukum, G. (1997). Implementation of dynamic strength models into 2D hydrocodes: Applications for atmospheric breakup and impact cratering. *International Journal of Impact Engineering*, *20*(1), 411–430. doi: https://doi.org/10.1016/S0734-743X(97)87511-2

Kasting, J. F. (1988). Runaway and moist greenhouse atmospheres and the evolution of Earth and Venus. *Icarus*, *74*(3), 472–494. doi: https://doi.org/10.1016/0019-1035(88)90116-9

Koll, D. D., & Cronin, T. W. (2018). Earth's outgoing longwave radiation linear due to h2o greenhouse effect. *Proceedings of the National Academy of Sciences*, *115*(41), 10293–10298.

Kring, D. A. (2007). The Chicxulub impact event and its environmental consequences at the Cretaceous–Tertiary boundary. *Palaeogeography, Palaeoclimatology, Palaeoecology*, *255*(1), 4-21. doi: https://doi.org/10.1016/j.palaeo.2007.02.037

Leconte, J., Forget, F., Charnay, B., Wordsworth, R., & Pottier, A. (2013). Increased insolation threshold for runaway greenhouse processes on Earth like planets. *Nature*, *504*(7479), 268–271. doi: https://doi.org/10.1038/nature12827

Le Hir, G., Donnadieu, Y., Krinner, G., & Ramstein, G. (2010). Toward the snowball earth deglaciation.... *Climate dynamics*, *35*, 285–297. doi: https://doi.org/10.1007/s00382-010
July 12, 2024, 5:54pm

-0748-8

Melosh, H. J., Ryan, E. V., & Asphaug, E. (1992). Dynamic fragmentation in impacts: Hydrocode simulation of laboratory impacts. *Journal of Geophysical Research: Planets*, *97*(E9), 14735–14759. doi: https://doi.org/10.1029/92JE01632

Pierrehumbert, R. T., Abbot, D. S., Voigt, A., & Koll, D. (2011). Climate of the Neoproterozoic. *Annual Review of Earth and Planetary Sciences*, *39*, 417–460. doi: https://doi.org/10.1146/annurev-earth-040809-152447

Spiegel, D. S., Menou, K., & Scharf, C. A. (2008). Habitable Climates. *The Astrophysical Journal*, *681*, 1609–1623. doi: https://doi.org/10.1086/588089

Thompson, S. L., & Lauson, H. S. (1974). *Improvements in the CHART D radiation-hydrodynamic code III: revised analytic equations of state* (Tech. Rep. No. SC-RR-71-0714). Sandia Laboratories.

Wagner, T. J., & Eisenman, I. (2015). How climate model complexity influences sea ice stability. *Journal of Climate*, *28*(10), 3998–4014.

Warren, S. G. (1982). Optical properties of snow. *Reviews of Geophysics*, *20*(1), 67–89. doi: https://doi.org/10.1098/rsta.2018.0161

Williams, D. M., Kasting, J. F., & Frakes, L. A. (1998). Low-latitude glaciation and rapid changes in the earth's obliquity explained by obliquity–oblateness feedback. *Nature*, *396*(6710), 453–455. doi: https://doi.org/10.1038/24845

Wünnemann, K., Collins, G. S., & Melosh, H. J. (2006). A strain-based porosity model for use in hydrocode simulations of impacts and implications for transient crater growth in porous targets. *Icarus*, *180*(2), 514–527. doi: https://doi.org/10.1016/j.icarus.2005.10.013

Wünnemann, K., Collins, G. S., & Osinski, G. R. (2008). Numerical modelling of impact melt production in porous rocks. *Earth and Planetary Science Letters*, *269*(3), 530–539. doi: https://doi.org/10.1016/j.epsl.2008.03.007




%
%
%
%
%

\end{document}